\documentclass[reprint,prl,amsmath,amssymb,aps,floatfix,superscriptaddress]{revtex4-2}
\usepackage{graphicx,booktabs,bm,microtype}
\usepackage[colorlinks=true,allcolors=blue]{hyperref}
\newcommand{\ket}[1]{\lvert#1\rangle}
\newcommand{\bra}[1]{\langle#1\rvert}
\newcommand{\ue}{\mu\mathrm{eV}}
\begin{document}
\title{Entanglement swapping by self-induced Larmor precession}
\author{Junhui Cao}
\affiliation{Abrikosov Center for Theoretical Physics, Moscow Center for Advanced Studies, Moscow 141701, Russia}
\author{Alexey Kavokin}
\affiliation{Russian Quantum Center, Skolkovo, Moscow Region 121205, Russia}
\affiliation{Abrikosov Center for Theoretical Physics, Moscow Center for Advanced Studies, Moscow 141701, Russia}
\affiliation{Department of Physics, St. Petersburg State University, University Embankment, 7/9, St. Petersburg 199034, Russia}
\affiliation{School of Science, Westlake University, 18 Shilongshan Road, Hangzhou 310024, Zhejiang Province, China}

\begin{abstract}
We propose an entanglement-swapping scheme based on self-induced Larmor precession in a polariton waveguide microcavity. Spin-dependent polariton interactions generate an effective field without an applied magnetic field. With one photon in each mode, the target precesses in opposite directions for the two circular control polarizations. A quantum superposition of these rotations entangles the polarization qubits. Local polarization rotations and detection convert this evolution into a complete Bell measurement, establishing entanglement between remote quantum memories. We find that spatial phase variation and spin dephasing select a finite optimal interaction range, and determine the resulting swapping fidelity, success probability, and entanglement-distribution rate. We argue that low losses ad high fidelity of the entangling operation make polariton Bell nodes highly promising for application in quantum repeaters.
\end{abstract}
\maketitle

Entanglement swapping couples two quantum links by a joint measurement on one particle from each link, leaving the remote particles entangled \cite{Zukowski1993}. Quantum repeaters use this operation to extend entanglement beyond the distance accessible by direct transmission \cite{Briegel1998,Sangouard2011,Azuma2023}. For polarization qubits, the conventional balanced-beam-splitter analyzer distinguishes two of the four Bell states, with intrinsic success probability $1/2$ \cite{Calsamiglia2001}. Auxiliary photons can increase this probability using larger linear optical networks \cite{Grice2011,Ewert2014}. An entangling interaction combined with local rotations can map the Bell basis onto a product basis for separate polarization readout. The required controlled-phase operation is also a basic quantum-computing gate. In a quantum repeater, resolving all four outcomes increases the swapping probability, while photon loss and memory dephasing reduce the gain.

Exciton--polaritons combine spin-dependent excitonic interactions with optical propagation and polarization readout \cite{cao2026emergent,cao2026polariton,mavrotsoupakis2025unveiling}. Electron--electron and hole--hole exchange contribute to exciton scattering \cite{Ciuti1998}, giving different polariton interaction strengths for parallel and antiparallel spins \cite{Renucci2005,Takemura2017}. A circular-polarization imbalance produces an effective field normal to the quantum-well plane. The in-plane projection of the polariton pseudospin precesses around this field, rotating the linear polarization of the emitted light.

Self-induced Larmor precession has been observed in polarization dynamics and described in the mean-field limit \cite{Renucci2005,Krizhanovskii2006,read2009stochastic,levrat2010pinning,Sigurdsson2022}. Spin-noise studies have also examined the polarization fluctuations associated with this precession \cite{Glazov2013,Ryzhov2020}. Experiments have demonstrated single-polariton propagation, photon--polariton entanglement, and few-photon polarization rotation \cite{SuarezForero2020,Cuevas2018,Kuriakose2022,PRXQuantum.5.010339}. By applying the same mechanism with one excitation in each of two optical modes, the control polarization will exhibit the potential for quantum entanglement. Reversing its circular polarization will reverse the target precession, while a superposition of control states creates a coherent superposition of target rotations. Hence the entanglement can develop even when the mean circular polarization of the control is zero.

Here we consider two single photons resonantly coupled into distinguishable overlapping modes of a polariton waveguide microcavity. The spin-dependent phase accumulates during the propagation. Different parts of the wavepackets acquire different rotations, allowing the outgoing temporal field to carry information about spin parity. This coupling between polarization and temporal modes reduces the coherence of photon phase gates \cite{He2012,Brod2016}. In entanglement swapping, the remote quantum memories inherit the resulting dephasing. We find a finite range that maximizes the swapped fidelity. Broadening the interaction makes the phase more uniform but slows the Bell operation and increases spin dephasing. We derive the optimal condition and  fidelity, success probability, and distribution rate from the evolution of the even- and odd-parity wavepackets.

\begin{figure*}[t]
\centering\includegraphics[width=0.8\textwidth]{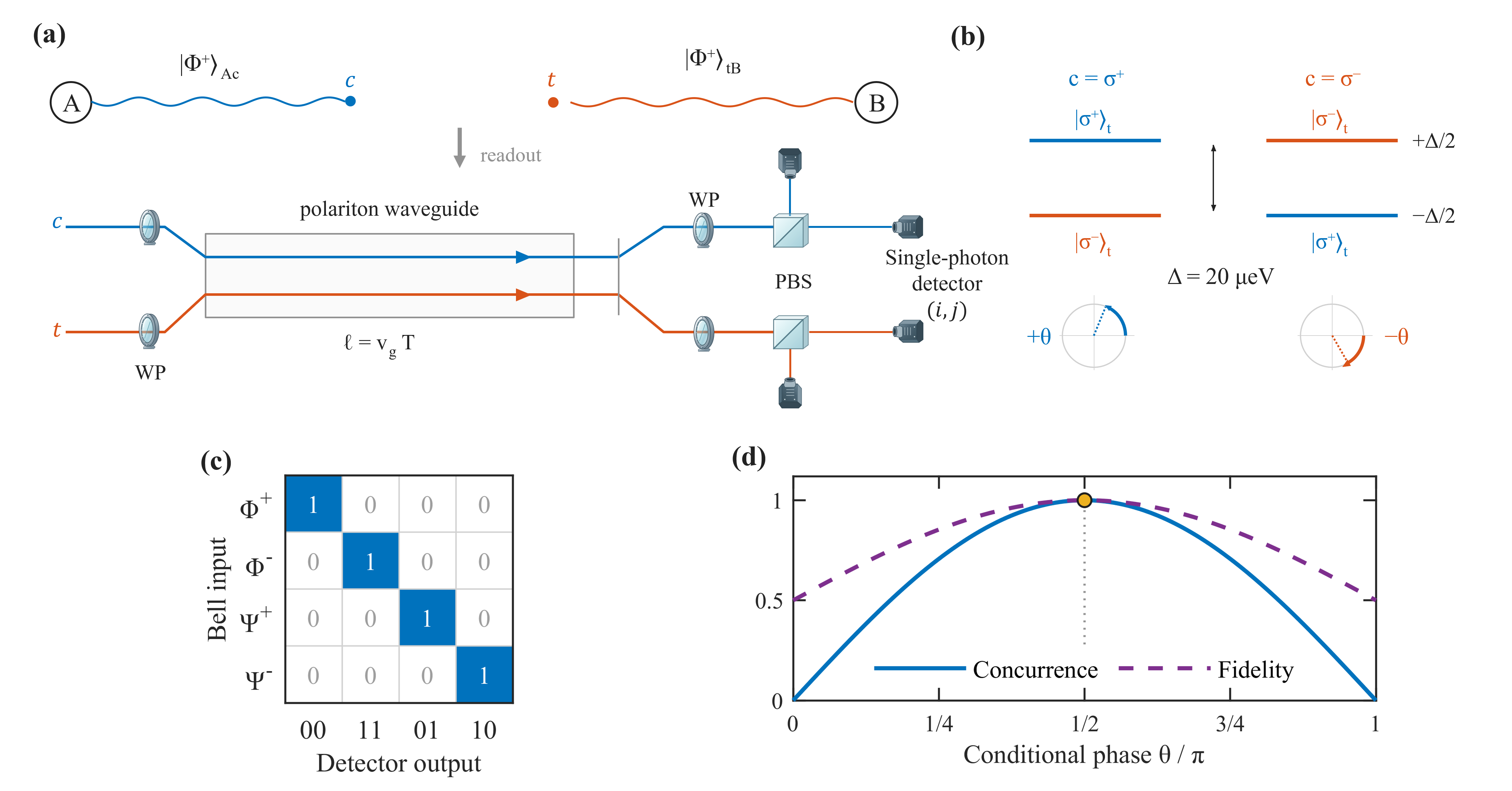}
\caption{Bell analyzer based on self-induced Larmor precession. (a) Photons $c,t$, initially entangled with quantum memories $A,B$, occupy distinct, overlapping modes in a waveguide microcavity. Waveplates (WP), mode separation and polarization-resolved detection identify the measurement outcome for a memory Pauli correction. (b) The interaction-induced effective field changes sign when the control's circular polarization is reversed. (c) Bell-state detection probabilities after the local rotations at $\theta=\pi/2$. (d) Concurrence generated from horizontal input polarizations and swapped fidelity for ideal input links. The Bell-measurement time is $51.7$ ps at $\Delta=20\,\ue$.}
\label{fig:mechanism}
\end{figure*}

\paragraph*{Model.}
We consider a GaAs-based microcavity containing quantum wells and patterned into a waveguide microcavity, supporting both polarizations of the lower-polariton branch with long lifetime \cite{Rozas2024,Walker2015,Steger2015}. Input and output dielectric gratings couple light to and from the propagating modes with separation $\ell$. Grating injection and optical readout have been demonstrated in semiconductor polariton waveguides \cite{Walker2015,Walker2019,fieramosca2026fully}. The two single photons are injected resonantly into distinguishable modes $c,t$ with overlapping transverse profiles and matched group velocities. They propagate together between the gratings and emerge into separate optical channels for polarization analysis. The interaction time is $T=\ell/v_{\rm gr}$, where $v_{\rm gr}=\partial\omega_{\rm LP}/\partial k$ is the group velocity. At $v_{\rm gr}=1.67\,\mu$m/ps, the reference time $51.7$ ps corresponds to $\ell=86.3\,\mu$m.

\paragraph*{Self-induced spin evolution.}
The circular polarizations define $\ket0=\ket{\sigma^+}$ and $\ket1=\ket{\sigma^-}$. Let $U_\parallel$ and $U_\perp$ be the interaction energies of parallel- and antiparallel-spin pairs in the two selected modes. We keep the diagonal interaction and neglect polarization exchange between the modes. For one excitation in each mode, removing their common energy shift gives
\begin{equation}
 H_{\rm int}=\frac{\Delta}{2}Z_cZ_t,\qquad
 U(T)=e^{-i\theta Z_cZ_t/2},\quad \theta=\frac{\Delta T}{\hbar}.
 \label{eq:H}
\end{equation}
Here $\Delta=U_\parallel-U_\perp$ and $Z$ is the Pauli operator. For a circular control with $Z_c=s=\pm1$, the target's Poincar\'e azimuth advances by $s\theta$ and its linear-polarization axis by $s\theta/2$. A linearly polarized control coherently correlates its two circular components with these opposite rotations. $\langle Z_c\rangle=0$ means the entangling evolution is discarded. Acting on two horizontal polarizations, $U$ gives concurrence $C=|\sin\theta|$ \cite{Wootters1998}. At $\theta=\pi/2$, the two conditional target states are orthogonal and the evolution is locally equivalent to a controlled-$Z$ gate. The first maximally entangling time is $T_*=\pi\hbar/(2\Delta)=51.7$ ps for the energy splitting $\Delta=20\,\ue$.

\paragraph*{Bell states and remote entanglement.}
Consider ideal elementary links $\ket{\Phi^+}_{Ac}$ and $\ket{\Phi^+}_{tB}$, where $A,B$ are remote memory qubits and $\ket{\Phi^+}=(\ket{00}+\ket{11})/\sqrt2$. Their joint state can be written as
\begin{equation}
 \begin{aligned}
 \ket{\Psi_0}&=\ket{\Phi^+}_{Ac}\ket{\Phi^+}_{tB}\\
 &=\frac12\sum_{\beta=\Phi^\pm,\Psi^\pm}\ket{\beta}_{AB}\ket{\beta}_{ct}.
 \end{aligned}
 \label{eq:links}
\end{equation}
An optical Bell projection prepares the corresponding memory Bell state. Local waveplates convert the conditional evolution into the analyzer $\mathcal B=(H\otimes H)U(H\otimes H)(S\otimes I)$, where $H=(X+Z)/\sqrt2$ and $S=\operatorname{diag}(1,i)$. At $T_*$, circular-polarization outcomes $00,01,10,11$ identify $\Phi^+,\Psi^+,\Psi^-,\Phi^-$, respectively. Applying $I,X,XZ,Z$ to memory $A$ makes every ideal outcome yield $\Phi^+$.

Let $E_m$ be the optical measurement effect associated with outcome $m$. For the input in Eq.~(\ref{eq:links}),
\begin{equation}
 \widetilde\rho_{AB}^{(m)}=\frac{E_m^{T}}4,\qquad
 p_m=\frac{\operatorname{Tr}E_m}{4}.
 \label{eq:effect}
\end{equation}
The transpose is taken in the circular product basis. Coherent evolution gives $E_m=\mathcal B^\dagger\ket m\bra m\mathcal B$. For example, $E_{00}=\ket{\chi}\bra{\chi}$ with $\ket{\chi}=\cos(\theta/2)\ket{00}+\sin(\theta/2)\ket{11}$. The measurement remains projective while its basis changes from product states at $T=0$ to Bell states at $T_*$. Since the optical input is $I/4$, each outcome has probability $1/4$ at every angle, which gives the heralded memory state for noisy measurement effects \cite{SM}.

\begin{figure*}[t]
\centering\includegraphics[width=0.8\textwidth]{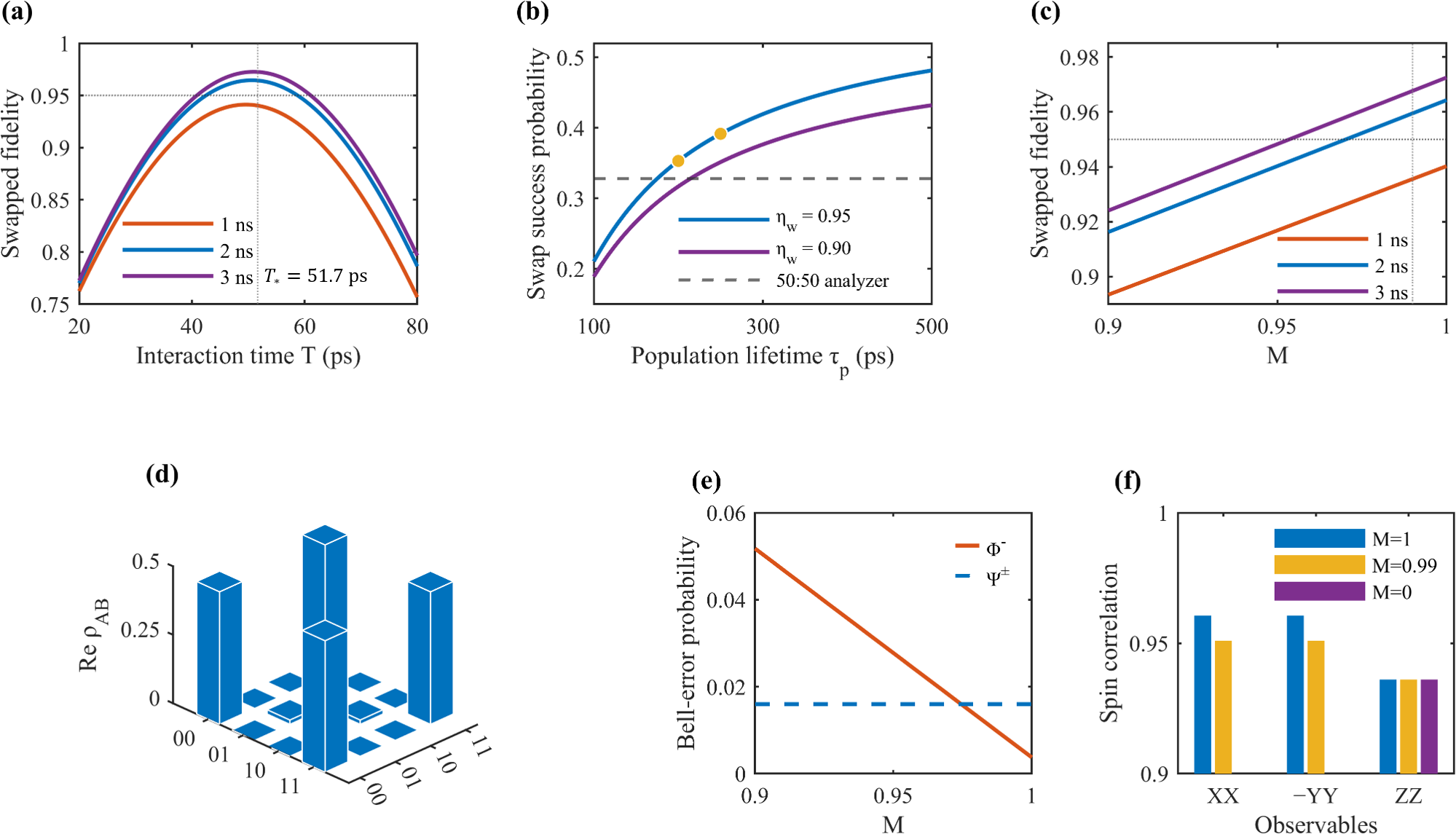}
\caption{Quantum coherence and heralded pair survival. (a) Swapped fidelity during the $M=1$ evolution for three spin-coherence times. The dashed line indicates $F_{\rm swap}=0.9501$. (b) Success probability at $T_*$ versus population lifetime. Markers show $200$ and $250$ ps at $\eta_w=0.95$. Dashed line shows the beam-splitter $P_{\rm swap}^{\rm BS}=0.328$. (c) Fidelity versus output-mode overlap at $T_*$. The horizontal dashed line indicates $F_{\rm swap}=0.9501$ while vertical one shows where $M=0.99$. (d) Real part of the corrected memory density matrix at $M=1$, $T_\phi=2$ ns. (e) Bell-error populations as $M$ varies. (f) Spin correlations for $M=1,0.99,0$, illustrating the loss of transverse coherence at fixed $ZZ$. Panels (c--f) use $T=T_*$ and (d--f) use $T_\phi=2$ ns. All results use $\Delta=20\,\ue$, $b=0.001$.}
\label{fig:simple}
\end{figure*}

\paragraph*{Coherence of the swapped state.}
The spin dynamics separates into even and odd parity parts, $\Pi_{e,o}=(I\pm Z_cZ_t)/2$. If they leave temporal states $u_e,u_o$, tracing those modes multiplies the coherence by $\Gamma=\langle u_o|u_e\rangle=M e^{-i\phi}$. For the uniform interaction, $M=1$ and $\phi=\theta$. The modulus $M$ measures the coherence after tracing the temporal field. Each spin also undergoes independent dephasing with $\lambda=e^{-T/T_\phi}$. For two Werner input links of fidelity $F_0$ and an unpolarized background fraction $b$ among accepted events, we define $A=(1-b)[(4F_0-1)/3]^2$. Averaging the memory states after the outcome-dependent Pauli corrections gives
\begin{equation}
 \begin{aligned}
 \rho_{AB}=\frac14\big[&I+A\lambda^2Z_AZ_B\\
 &+A\lambda s(X_AX_B-Y_AY_B)\big],
 \end{aligned}
 \label{eq:rho}
\end{equation}
where $s=-\operatorname{Im}\Gamma$. This Bell-diagonal state has unequal transverse and longitudinal spin correlations. Its fidelity with the target $\Phi^+$ is
\begin{equation}
 \begin{aligned}
 F_{\rm swap}&=\frac{1+\langle XX\rangle-\langle YY\rangle+\langle ZZ\rangle}{4}\\
 &=\frac14+\frac A4\left(\lambda^2+2\lambda s\right).
 \end{aligned}
 \label{eq:F}
\end{equation}
Here $XX$ abbreviates $X_A X_B$, and similarly for $YY,ZZ$. The intrinsic node fidelity $F_n$ follows by setting $A=1$, removing source and background errors. The $ZZ$ term describes population correlations and the $XX-YY$ term contains the coherence between $\ket{00}$ and $\ket{11}$. At $T_*$, decreasing $M$ transfers weight from $\Phi^+$ to $\Phi^-$ while leaving the $\Psi^\pm$ amplitudes unchanged, as shown in Fig.~\ref{fig:simple}. Complete loss of this coherence leaves a separable, classically correlated state even when $\langle ZZ\rangle$ remains large.

Teleportation through this pair produces a Pauli channel whose probabilities are the Bell populations \cite{Bowen2001}. The $\Phi^-$ component gives a phase flip, while $\Psi^+$ and $\Psi^-$ give $X$ and $Y$ errors. At fixed conditional phase and other noise, mode overlap $M$ is equivalent to an additional $Z$-flip channel with probability $(1-M)/2$. $M=0.99$ therefore adds $0.5\%$ phase-flip noise. Ideal local teleportation operations give average state-transfer fidelity $(2F+1)/3$ for a resource of Bell fidelity $F$ \cite{Horodecki1999}, or at least $0.9667$ at the chosen repeater threshold. With $T_\phi=2$ ns, $F_0=0.995$, and $b=0.001$, the uniform-interaction model at $T_*$ gives $F_{\rm swap}=0.96430$ and off-diagonal coherence $\rho_{00,11}=0.48029$.

\begin{figure*}[t]
\centering\includegraphics[width=0.8\textwidth]{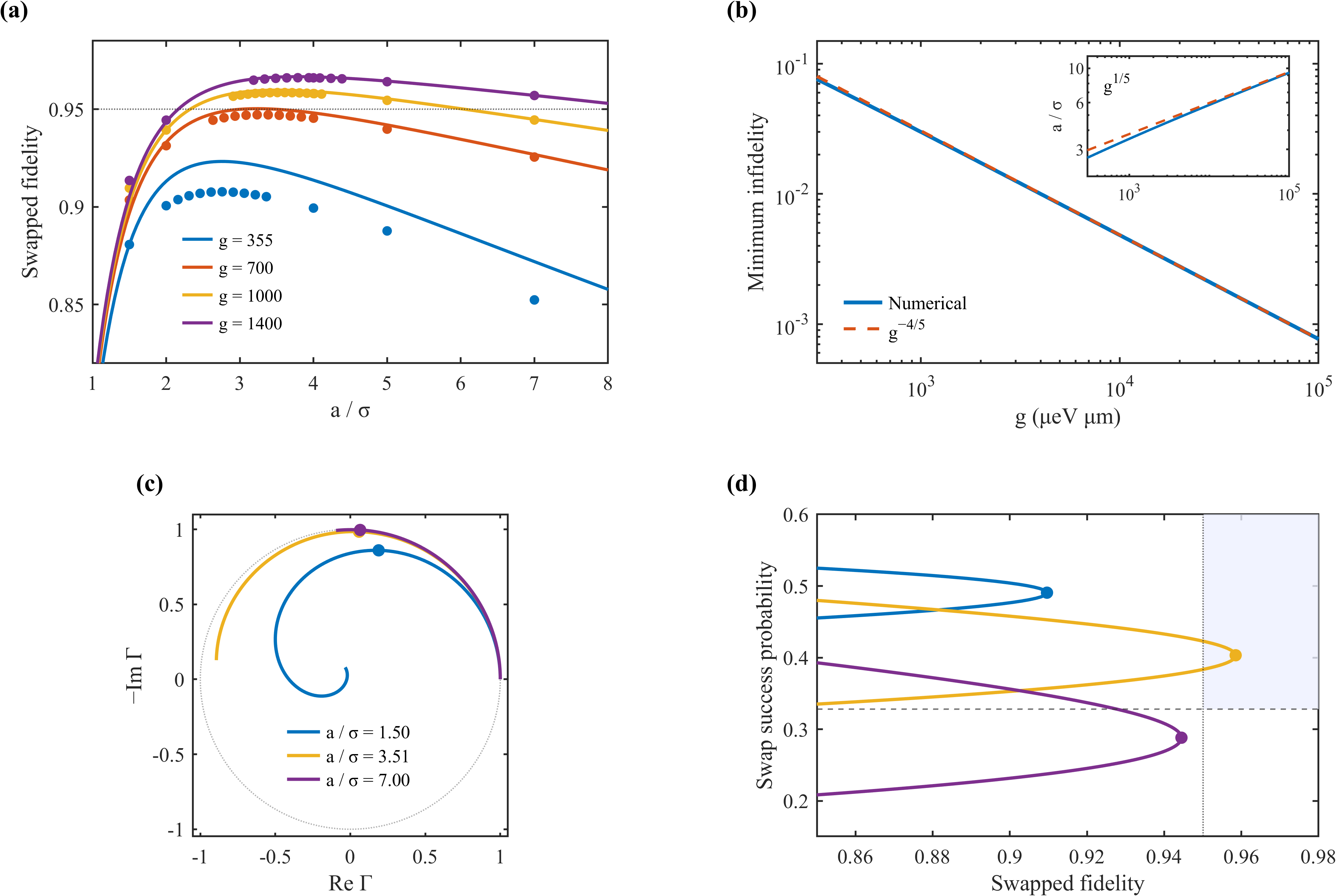}
\caption{Finite optimal range of the spin-dependent polariton interaction. (a) Swapped fidelity at fixed integrated coupling $g$ (in $\ue\,\mu$m). Lines indicate dispersionless approximation results at $T_q$. Points are maximum fidelity from dispersive propagation over $0\leq T\leq200$ ps. Dotted line marks $F_{\rm swap}=0.9501$. (b) Intrinsic node infidelity optimized over range at $T_q$, compared with the $g^{-4/5}$ law. The inset shows the optimal range and its $g^{1/5}$ asymptote. The high-coupling extension tests the asymptotic law. (c) Complex parity overlap for $g=1000\,\ue\,\mu$m and three ranges, over $0\leq T\leq100$ ps. Circles mark the maximum local fidelity. (d) Fidelity versus success probability for $0\leq T\leq200$ ps at $\tau_p=250$ ps. Shading indicates $F_{\rm swap}\geq0.9501$ and swapping probability above the beam-splitter reference. The horizontal dashed line shows $P_{\rm swap}^{\rm BS}=0.328$. All panels use $\sigma=5\,\mu$m and $T_\phi=2$ ns. The dispersive calculation uses $\kappa=0.8356\,\mu$m$^2$/ps.}
\label{fig:range}
\end{figure*}

\paragraph*{Optimal range.}
Let each incident photon have a Gaussian intensity envelope with width $\sigma$. The relative longitudinal coordinate $r$ then has variance $2\sigma^2$. We model the difference between the even- and odd-parity interaction energies by
\begin{equation}
 \delta(r)=\frac{g}{\sqrt{2\pi}a}e^{-r^2/(2a^2)},\qquad
 \overline\Delta=\frac{g}{\sqrt{2\pi}\sqrt{a^2+2\sigma^2}}.
 \label{eq:gaussian}
\end{equation}
Here $g=\int\delta(r)dr$ is the integrated coupling, $a$ is the longitudinal interaction range, and $\overline\Delta$ is the mean splitting at the input. Increasing $a$ at fixed $g$ makes the phase more uniform but lowers $\overline\Delta$, while keeping the mean splitting fixed would require an increasing $g$.

The two temporal amplitudes evolve with $H_j=-\hbar\kappa\partial_r^2+V_j(r)$, where $V_e-V_o=\delta$ and we use $V_o=-0.1V_e$. We evaluate $\Gamma(T)=\langle u_o(T)|u_e(T)\rangle$ over the full outgoing amplitudes, collecting all arrival times. In the dispersionless approximation, their overlap reduces to
\begin{equation}
 \Gamma(T)=\int\!dr\,\frac{e^{-r^2/(4\sigma^2)}}{2\sqrt\pi\sigma}
 e^{-i\delta(r)T/\hbar}.
 \label{eq:gaussgamma}
\end{equation}
At fixed $a$ and $\sigma$, this overlap depends on $gT$, as increasing $g$ shortens the time required to accumulate the target phase.

For a broad interaction with a Gaussian spatial profile, the relative variance of the local splitting is $\operatorname{Var}(\delta)/\overline\Delta^{\,2}\approx2(\sigma/a)^4$. Expanding Eq.~(\ref{eq:gaussgamma}) at the time $T_q=\pi\hbar/(2\overline\Delta)$, corresponding to a mean conditional phase of $\pi/2$, gives
\begin{equation}
 1-F_n\approx\frac{T_q}{T_\phi}+\frac{\pi^2\sigma^4}{8a^4}
 \approx\zeta q+\frac{\pi^2}{8q^4},
 \label{eq:balance}
\end{equation}
where $q=a/\sigma$ and $\zeta=\pi\hbar\sqrt{2\pi}\sigma/(2gT_\phi)$. The first term is spin dephasing during the operation while the second is the loss of polarization coherence through phase variation. Their balance gives the optimal range $q_{\rm opt}$ and minimum node infidelity $1-F_n$,
\begin{equation}
 \begin{aligned}
 q_{\rm opt}&\approx\left(\frac{\pi^2}{2\zeta}\right)^{1/5},\\
 (1-F_n)_{\min}&\approx\frac54\left(\frac{\pi^2}{2}\right)^{1/5}\zeta^{4/5}.
 \end{aligned}
 \label{eq:rangeopt}
\end{equation}
\begin{figure*}[t]
\centering\includegraphics[width=0.8\textwidth]{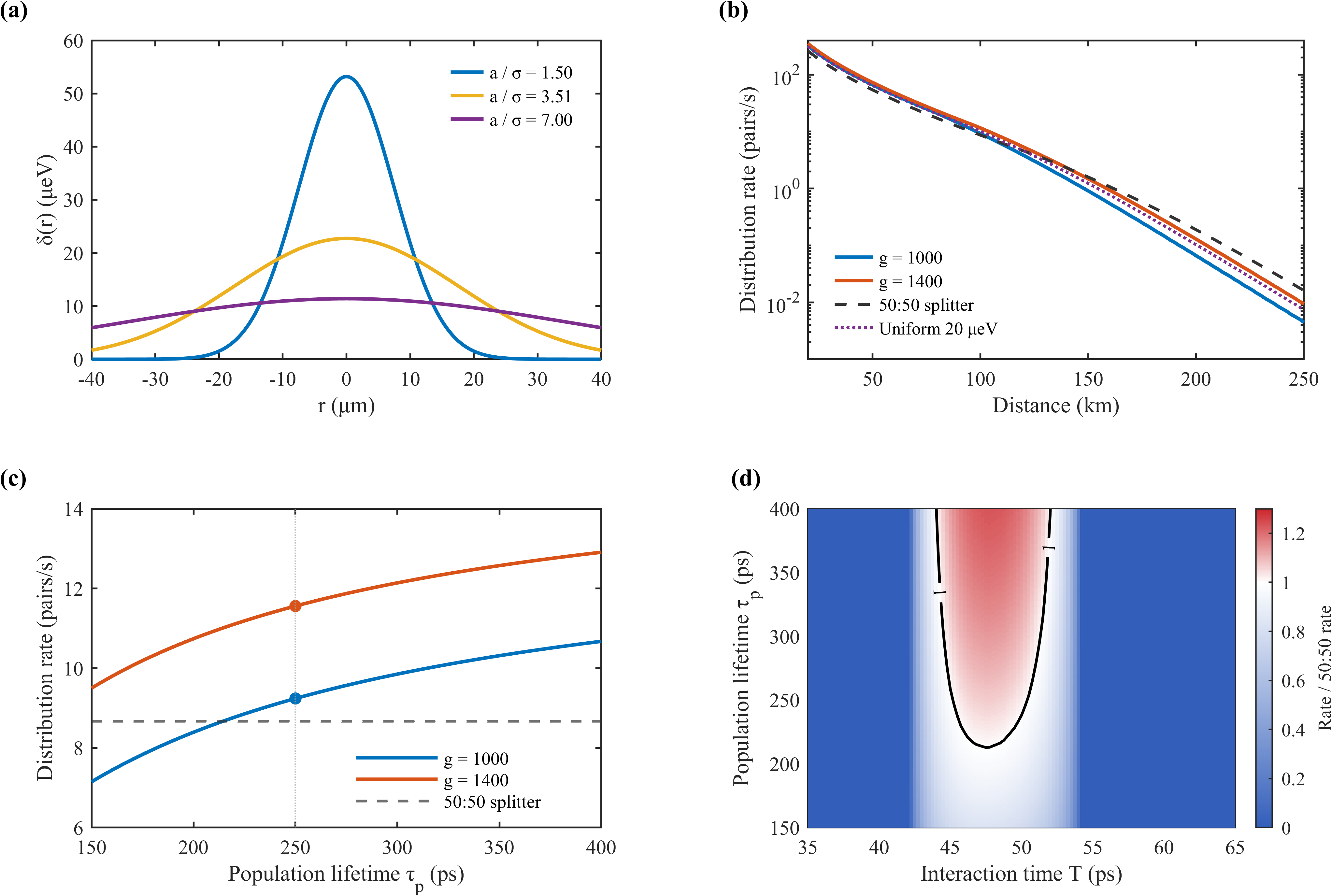}
\caption{Entanglement distribution. (a) Interaction-induced splitting profiles at fixed $g=1000\,\ue\,\mu$m and three ranges. (b) Distribution rate versus total distance, compared with the balanced-beam-splitter analyzer (gray dashed) and the uniform-interaction $20\,\ue$ reference at $T_*$ and $\tau_p=250$ ps (purple dotted). (c) Rate versus population lifetime at $100$ km. The horizontal dashed line is the beam-splitter reference. (d) Rate relative to the beam-splitter analyzer versus interaction time and population lifetime for $g=1000\,\ue\,\mu$m$,\ a/\sigma=3.510$, and $L=100$ km. The solid curve marks equal rates. In (b--d), only the memory waiting-time cutoff is optimized at each point, subject to the ensemble-mean fidelity $F_{\rm out}\geq0.9501$.}
\label{fig:rates}
\end{figure*}

At fixed packet width and weak Markovian spin dephasing, the optimal range grows as $a_{\rm opt}\propto g^{1/5}$ and the intrinsic error falls as $g^{-4/5}$. Numerical evaluation of the phase integral gives an exponent of $-0.799$ in the high-coupling regime and dispersive propagation at the corresponding times $T_q$ gives $-0.800$ \cite{SM}. At smaller couplings, the full dynamics shifts the fidelity and optimum, as shown in Fig.~\ref{fig:range}(a). For $g=1000\,\ue\,\mu$m and $\sigma=5\,\mu$m, the best point in the range scan is $a=17.55\,\mu$m, with peak local fidelity at $T=48.1$ ps. The mean splitting at the input is $21.08\,\ue$. Propagating both parity branches gives $\Gamma=0.055-0.984i$, hence $M=0.986$ and $F_{\rm swap}=0.959$. The overlap trajectories show the loss of coherence for a narrower interaction and the slower precession for a broader one (Fig.~\ref{fig:range}(c)).

\paragraph*{Swapping probability.}
Given two prepared elementary links, a successful attempt is heralded by detecting one photon from each of the two output modes $c$ and $t$. All four polarization outcomes are accepted and determine the Pauli correction. A missing detection means a failure. Polarization errors in accepted events will reduce the swapped-state fidelity. Spin-independent photon loss reduces the swapping probability while preserving the conditional state in Eq.~(\ref{eq:rho}). For population lifetime $\tau_p$, combined input--output coupling efficiency $\eta_w$ per photon, memory-readout efficiency $\eta_r$, and detector efficiency $\eta_d$, we have
\begin{equation}
 P_{\rm swap}=(\eta_r\eta_d)^2\eta_w^2e^{-2T/\tau_p}.
 \label{eq:P}
\end{equation}
We use $\eta_w=0.95$ and $\eta_r=\eta_d=0.9$. The conventional beam-splitter analyzer has $P_{\rm BS}=(\eta_r\eta_d)^2/2=0.328$. $\tau_p=250$ ps gives $P_{\rm swap}=0.403$ (Fig.~\ref{fig:range}(d)).

\paragraph*{Entanglement distribution.}
We then connect two equal elementary links across total distance $L$. Each succeeds with probability $p=0.1\times10^{-0.2L/20}$ per attempt, using $0.2$ dB/km telecom attenuation and including conversion losses in the preparation or interface efficiencies. Once one elementary link is established, its entangled state is stored in quantum memories while generation attempts continue on the other link. Memory decoherence during this waiting time reduces the fidelity of the stored entanglement. We allow at most $D$ further attempts before resetting both links. A failed entanglement swap will also reset both. Memory noise is isotropic depolarization with a stored-link visibility lifetime of $1$ s, contracting all three correlations in Eq.~(\ref{eq:rho}). We calculate the delivered fidelity and cycle duration from the link-waiting distribution, including herald communication \cite{SM}.

Figure~\ref{fig:rates} shows that at $100$ km, the parameter set gives $9.24$ pairs/s for $\tau_p=250$ ps, compared with $8.67$ pairs/s for the conventional analyzer. The delivered ensemble has fidelity $0.95015$. Reducing the lifetime to $200$ ps leaves the local fidelity unchanged but lowers the rate to $8.39$ pairs/s. At this fixed interaction time and range, exceeding the beam-splitter rate requires a lifetime above $214.5$ ps. Increasing the integrated coupling to $1400\,\ue\,\mu$m gives a best scanned range $a/\sigma=3.787$ and interaction time $36.8$ ps. The mean splitting is $27.63\,\ue$, with $F_{\rm swap}=0.966$, $P_{\rm swap}=0.441$, and $R=11.56$ pairs/s at $100$ km and $\tau_p=250$ ps. At the smaller fixed coupling $g=354.63\,\ue\,\mu$m, broadening the interaction gives a best scanned fidelity of $0.908$. As distance increases, memory decoherence during the waiting time consumes the initial fidelity margin and forces more frequent resets (Fig.~\ref{fig:rates}(b)). Figure~\ref{fig:rates}(d) shows where the interacting node exceeds the beam-splitter rate, using the propagated $\Gamma(T)$ and optimizing the cutoff at each point. Stopping the interaction earlier sacrifices some polarization coherence but lets more photon pairs survive. Circular control inputs and target Stokes measurements determine the conditional rotation. Control polarization superpositions and joint polarization measurements probe the coherence between the two precession amplitudes. Measurements of $XX,YY,ZZ$ on the remote memories then give the swapped fidelity through Eq.~(\ref{eq:F}).  At velocity $1.67\,\mu$m/ps, the $48.1$ ps operation corresponds to $80.3\,\mu$m of propagation.

\paragraph*{Conclusion.}
Self-induced Larmor precession becomes a conditional quantum operation when the control is a single polarization qubit. The resulting two-polariton phase enables a complete Bell measurement, with the waveguide length setting the operation time. At fixed integrated strength, phase variation and spin dephasing select a finite optimal range, with intrinsic error scaling asymptotically as $g^{-4/5}$. Coherent photon--polariton conversion allows the Bell operation to act on propagating polarization qubits without an additional storage and retrieval step inside the analyzer. Together with the high-fidelity entanglement swapping through self-induced Larmor precession, this makes polariton Bell nodes promising for quantum repeaters.

\paragraph*{Acknowledgments.---}
AK acknowledges support from Saint Petersburg State University (Research Grant No. 125022803069-4) and from the Innovation Program for Quantum Science and Technology (No. 2021ZD0302704). 

\bibliographystyle{apsrev4-2}
\bibliography{references}
\end{document}


\maketitle

\section{Self-induced Larmor precession and Bell measurement}
Spin-dependent excitonic interactions distinguish parallel- and antiparallel-spin polariton pairs while carrier exchange contributes to this difference \cite{Ciuti1998,Renucci2005,Takemura2017}. In a polariton fluid, a circular-polarization imbalance produces an interaction-induced splitting proportional to $(\alpha_\parallel-\alpha_\perp)(n_+-n_-)$, where $\alpha_\parallel,\alpha_\perp$ are the corresponding interaction coefficients and $n_\pm$ are the circular-component densities. This splitting acts as an effective field normal to the quantum wells, causing the in-plane Stokes vector to precess without an applied magnetic field.

For two distinguishable modes $c,t$, let $n_{j\pm}$ be the occupation operators of their circular components. Neglecting polarization exchange between the modes and scattering into other modes, the diagonal interaction is
\begin{equation}
 H_{ct}=U_\parallel(n_{c+}n_{t+}+n_{c-}n_{t-})
       +U_\perp(n_{c+}n_{t-}+n_{c-}n_{t+}).
 \label{eq:spininteraction}
\end{equation}
The energies $U_\parallel,U_\perp$ include projection onto the two spatial modes. With $N_c=N_t=1$, the circular basis is $\ket0=\ket{\sigma^+}$, $\ket1=\ket{\sigma^-}$, and
\begin{equation}
 H_{ct}=\frac{U_\parallel+U_\perp}{2}I+
 \frac{\Delta}{2}Z_cZ_t,\qquad \Delta=U_\parallel-U_\perp.
\end{equation}
Removing the common energy shift gives $U=\exp(-i\theta Z_cZ_t/2)$, with $\theta=\Delta T/\hbar$. For a circularly polarized control photon labeled by $s=\pm1$, the target splitting is $s\Delta$. The target's azimuthal angle on the Poincar\'e sphere changes by $s\theta$, corresponding to a rotation of its linear-polarization axis by $s\theta/2$. For two horizontal input polarizations, $\ket{+_x}=(\ket0+\ket1)/\sqrt2$,
\begin{equation}
 U\ket{+_x,+_x}=\frac{e^{-i\theta/2}(\ket{00}+\ket{11})+
 e^{i\theta/2}(\ket{01}+\ket{10})}{2}.
\end{equation}
The concurrence is $C=|\sin\theta|$. The first maximally entangling time is $T_*=\pi\hbar/(2\Delta)=51.7$ ps for the reference conditional splitting $\Delta=20\,\ue$ induced by one control excitation. Although this input has $\langle Z_c\rangle=0$, its two control components remain coherently correlated with opposite target rotations.

Local waveplates implement $H=(X+Z)/\sqrt2$ and $S=\operatorname{diag}(1,i)$, giving the analyzer $\mathcal B=(H\otimes H)U(H\otimes H)(S\otimes I)$. At $T_*$,
\begin{equation}
 \ket{\Phi^+}\mapsto\ket{00},\qquad \ket{\Psi^+}\mapsto\ket{01},\qquad
 \ket{\Psi^-}\mapsto-i\ket{10},\qquad \ket{\Phi^-}\mapsto-i\ket{11}.
 \label{eq:bellmap}
\end{equation}
Here $\ket{\Phi^\pm}=(\ket{00}\pm\ket{11})/\sqrt2$ and $\ket{\Psi^\pm}=(\ket{01}\pm\ket{10})/\sqrt2$. For initial links $\ket{\Phi^+}_{Ac}$ and $\ket{\Phi^+}_{tB}$, the records $00,01,10,11$ require corrections $I,X,XZ,Z$ on memory $A$, respectively.

\section{Measurement effects and the heralded memory state}
The initial state of the two links can be written as
\begin{equation}
 \ket{\Psi_0}=\frac12\sum_{i,j=0}^1
 \ket{ij}_{AB}\ket{ij}_{ct}
 =\frac12\sum_{\beta=\Phi^\pm,\Psi^\pm}\ket\beta_{AB}\ket\beta_{ct}.
\end{equation}
Let $V=(H\otimes H)(S\otimes I)$, $W=H\otimes H$. The measurement effect for outcome $m$ is
\begin{equation}
 E_m=V^\dagger
 (W^\dagger\ket m\bra m W)V,\qquad \sum_m E_m=I.
 \label{eq:effects}
\end{equation}
For the unitary evolution $U$, this reduces to $E_m=\mathcal B^\dagger\ket m\bra m\mathcal B$. For example,
\begin{equation}
 E_{00}=\ket{\chi_{00}}\bra{\chi_{00}},\qquad
 \ket{\chi_{00}}=\cos(\theta/2)\ket{00}+\sin(\theta/2)\ket{11}.
\end{equation}
The measurement basis evolves from product states at $T=0$ to Bell states at $T_*$. The optical input $I/4$ is unchanged by the unital polarization channel.

The unnormalized memory state is $\widetilde\rho_{AB}^{(m)}=E_m^{T}/4$, with probability $p_m=\operatorname{Tr}E_m/4$; the transpose is taken in the circular product basis. Here each effect has trace one, so all four outcomes have probability $1/4$. With $Q_m=I,X,XZ,Z$ for $m=00,01,10,11$, the corrected ensemble is
\begin{equation}
 \rho_{AB,n}=\frac14\sum_m(Q_m\otimes I)E_m^{\mathsf T}
 (Q_m^\dagger\otimes I).
 \label{eq:conditionalensemble}
\end{equation}

Independent $Z$ dephasing of each polarization qubit has coherence $\lambda_j=e^{-T/T_{\phi,j}}$ and channel $\mathcal D_j(\rho)=(1+\lambda_j)\rho/2+(1-\lambda_j)Z_j\rho Z_j/2$. These errors commute with $U$. For ideal input links, the node fidelity is
\begin{equation}
 F_n=\frac{1+\lambda_c\lambda_t+
 (\lambda_c+\lambda_t)\sin\theta}{4},\qquad 0\leq\theta\leq\pi.
 \label{eq:Fn}
\end{equation}
Werner input links of fidelity $F_0$ have visibility $v_0=(4F_0-1)/3$. For an unpolarized background fraction $b$ among accepted events, define $A=(1-b)v_0^2$. The local swapping fidelity, before decoherence in the memories, is $F_{\rm swap}=1/4+A(F_n-1/4)$. For equal dephasing factors $\lambda_j=\lambda$ and parity-mode overlap $\Gamma$ defined below, set $s=-\operatorname{Im}\Gamma$. Equation~(\ref{eq:conditionalensemble}) gives
\begin{equation}
 \rho_{AB}=\frac14
 \begin{pmatrix}
 1+A\lambda^2&0&0&2A\lambda s\\
 0&1-A\lambda^2&0&0\\
 0&0&1-A\lambda^2&0\\
 2A\lambda s&0&0&1+A\lambda^2
 \end{pmatrix}.
 \label{eq:matrix}
\end{equation}
Its nonzero correlations are $\langle XX\rangle=-\langle YY\rangle=A\lambda s$ and $\langle ZZ\rangle=A\lambda^2$. Local Bloch terms in individual outcomes cancel upon averaging the Pauli-corrected states. The resulting Bell populations are
\begin{equation}
 p_{\Phi^\pm}=\frac{1+A\lambda^2\pm2A\lambda s}{4},\qquad
 p_{\Psi^\pm}=\frac{1-A\lambda^2}{4}.
 \label{eq:bellpop}
\end{equation}
Thus $F_{\rm swap}=p_{\Phi^+}$. For $s\geq0$, this is the largest Bell population, and the concurrence is $C=\max(0,2F_{\rm swap}-1)$ \cite{Wootters1998}. The negativity is $C/2$. The witness $I/2-\ket{\Phi^+}\bra{\Phi^+}$ has expectation $1/2-F_{\rm swap}$, so the three correlations also detect entanglement when $F_{\rm swap}>1/2$.

At $T_*=51.7$ ps, $T_\phi=2$ ns, $F_0=0.995$, and $b=0.001$, the populations in the order $\Phi^+,\Phi^-,\Psi^+,\Psi^-$ are $(0.9643,0.0037,0.0160,0.0160)$ for $M=1$ and $(0.9595,0.0085,0.0160,0.0160)$ for $M=0.99$. Reducing $M$ transfers population from $\Phi^+$ to $\Phi^-$ while preserving the $ZZ$ correlation. For $M=1$, the concurrence is $0.9286$.

\section{Photon survival and polarization coherence}
For population lifetime $\tau_p$, input--output coupling efficiency $\eta_w$ per photon, memory-readout efficiency $\eta_r$, and detector efficiency $\eta_d$, the swapping probability is $P_{\rm swap}=(\eta_r\eta_d)^2\eta_w^2e^{-2T/\tau_p}$. It is conditioned on both elementary links being ready. A herald requires one detected photon from each output mode $c,t$; all four polarization outcomes are accepted. Spin-independent loss changes this probability but preserves the normalized conditional state. We use $\eta_w=0.95$ and $\eta_r=\eta_d=0.9$. The conventional two-state beam-splitter analyzer has $P_{\rm BS}=(\eta_r\eta_d)^2/2=0.328$. At $T_*$, $P_{\rm swap}>P_{\rm BS}$ requires
\begin{equation}
 \tau_p>\frac{2T_*}{\ln(2\eta_w^2)},\qquad \eta_w>1/\sqrt2.
\end{equation}
For $\eta_w=0.95$, the lifetime threshold is $175.1$ ps. With photon loss alone, the earliest time to reach ideal-link fidelity $F$ is $T_{\min}=\hbar\arcsin(2F-1)/\Delta$ for $1/2\leq F\leq1$. The competition between phase accumulation and photon survival can favor an operation shorter than $T_*$.

For normalized even- and odd-parity output wavepackets $u_e,u_o$, define $\Gamma=\langle u_o|u_e\rangle=M e^{-i\phi}$ and $\Pi_{e,o}=(I\pm Z_cZ_t)/2$. Tracing over the temporal modes gives
\begin{equation}
 \mathcal E(\rho)=\Pi_e\rho\Pi_e+\Pi_o\rho\Pi_o+
 \Gamma\Pi_e\rho\Pi_o+\Gamma^*\Pi_o\rho\Pi_e.
\end{equation}
The $\sin\theta$ term in Eq.~(\ref{eq:Fn}) becomes $-\operatorname{Im}\Gamma=M\sin\phi$. Here $M$ is the overlap of the two outgoing parity branches. For the uniform interaction, $M=1$ and $\phi=\theta$.

Before memory decoherence, the overlap required at $\phi=\pi/2$ is
\begin{equation}
 M\geq\frac{4f_{\rm req}-1-\lambda^2}{2\lambda},\qquad
 f_{\rm req}=\frac14+
 \frac{F_{\min}-1/4}{(1-b)v_0^2}.
\end{equation}
At $T_*$, with $F_{\min}=0.9501$, $F_0=0.995$, $b=0.001$, and $T_\phi=2$ ns, the required overlap is approximately $0.9704$. For $M=1,0.98,0.95$, the swapping fidelities are $0.9643$, $0.9547$, and $0.9403$, respectively.

For a circularly polarized control and an initially horizontally polarized target, the target Stokes parameters are
\begin{equation}
 \langle X\rangle=\lambda_tM\cos\phi,\qquad
 \langle Y\rangle=\pm\lambda_tM\sin\phi,
\end{equation}
where the sign is set by the control polarization. A single-photon reference determines $\lambda_t$. Control polarization superpositions and joint polarization measurements probe the coherence of the conditional evolution and measurements of $XX,YY,ZZ$ on the remote memories give $F_{\rm swap}$.

\section{Entanglement distribution}
Each elementary link succeeds independently with probability $p$ per attempt of duration $\tau_0$. The first successful link is stored while the other continues its generation attempts, for at most $D$ additional attempts. A timeout or failed entanglement swap resets both links. With $q=1-p$, $w_0=p/(2-p)$, and $w_1=2q/(2-p)$, the probability that both links become ready and the mean number of preparation attempts per cycle are
\begin{equation}
 A_D=w_0+w_1(1-q^D),\qquad
 \bar n_D=\frac1{p(2-p)}+\frac{w_1}{p}(1-q^D).
\end{equation}
Memory noise is isotropic depolarization. If the stored link's Werner visibility decays with time constant $T_m$, set $\xi=e^{-\tau_0/T_m}$. The mean visibility factor for cycles with both links ready is
\begin{equation}
 C_D=\frac{w_0+w_1p\xi[1-(q\xi)^D]/(1-q\xi)}{A_D}.
\end{equation}
For unlimited waiting time, $A_\infty=1$, $\bar n_\infty=(3-2p)/[p(2-p)]$, and $C_\infty=w_0+w_1p\xi/(1-q\xi)$. For $D=0$ we have $C_0=1$. The average delivered fidelity and pair-distribution rate are
\begin{equation}
 F_{\rm out}=\frac14+(1-b)v_0^2 C_D e^{-t_{\rm comm}/T_m}
 (F_n-1/4),\qquad
 R=\frac{A_DP_{\rm swap}}{\tau_0\bar n_D+A_Dt_s}.
 \label{eq:renewal}
\end{equation}
The delivered state has the form of Eq.~(\ref{eq:matrix}) with $A$ replaced by $A C_D e^{-t_{\rm comm}/T_m}$. Here $t_{\rm comm}=L/(2c_f)$ is the herald communication delay across half the total distance, with fiber propagation speed $c_f=2\times10^5$ km/s. The swapping stage lasts $t_s=t_{\rm comm}+1\,\mu$s, including the local operation overhead. We use $\tau_0=t_s$, $T_m=1$ s, and $p=0.1\times10^{-0.2L/20}$ with $L$ in km. The attenuation is $0.2$ dB/km at telecom wavelengths. The input fidelity $F_0=0.995$ is defined when an elementary link is heralded, and $b=0.001$. We optimize $D$ subject to $F_{\rm out}\geq0.9501$, averaged over accepted storage times.

\begin{center}
\begin{tabular}{lrrrr}
\toprule
Uniform interaction at $100$ km & $T$ (ps)&$P_{\rm swap}$&$D$&$R$ (s$^{-1}$)\\
\midrule
$\tau_p=200$ ps, $T=T_*$ &51.7&0.353&263&9.10\\
$\tau_p=250$ ps, $T=T_*$ &51.7&0.392&263&10.09\\
$\tau_p=200$ ps, time optimized &48.7&0.364&243&9.32\\
$\tau_p=250$ ps, time optimized &49.0&0.400&252&10.28\\
Conventional two-state beam splitter &---&0.328&$\infty$&8.67\\
\bottomrule
\end{tabular}
\end{center}
The interacting rows use $\Delta=20\,\ue$, $M=1$, and $T_\phi=2$ ns. Calculations use unrounded operating times. At $200$ km, the time-optimized rates for $\tau_p=200$ and $250$ ps are $0.096$ and $0.106$ s$^{-1}$, compared with $0.188$ s$^{-1}$ for the beam splitter. At $100$ km and $T_*$, reducing $M$ to $0.98$ gives rates of $6.42$ and $7.12$ s$^{-1}$ for these two lifetimes.

With photon-number-resolving detection, the beam-splitter analyzer can also accept a fraction $x$ of ambiguous outcomes, each with Bell overlap $1/2$. Its intrinsic acceptance probability is $(1+x)/2$ and node fidelity is $(2+x)/[2(1+x)]$. Optimizing $x$ and $D$ under the same average-fidelity constraint gives $9.07$ pairs/s at $100$ km.

\section{Teleportation channel and communication errors}
With ideal local Bell measurement and feed-forward, the swapped Bell-diagonal state gives the teleportation channel \cite{Bowen2001}
\begin{equation}
 \mathcal T(\sigma)=p_{\Phi^+}\sigma+p_{\Psi^+}X\sigma X
 +p_{\Psi^-}Y\sigma Y+p_{\Phi^-}Z\sigma Z,
 \label{eq:teleportation}
\end{equation}
where $\sigma$ is the input-qubit density matrix. Set $G=A$ for the local swapped state and $G=A C_D e^{-t_{\rm comm}/T_m}$ for the delivered pair, and define $c_\perp=G\lambda M\sin\phi$, $c_z=G\lambda^2$. The input Bloch vector transforms as $(r_x,r_y,r_z)\mapsto(c_\perp r_x,c_\perp r_y,c_z r_z)$. For a pure input at polar angle $\vartheta$ in the circular basis,
\begin{equation}
 f_{\rm tel}(\vartheta)=\frac{1+c_\perp\sin^2\vartheta+c_z\cos^2\vartheta}{2},
 \qquad \overline f_{\rm tel}=\frac{2F_{\rm res}+1}{3}.
\end{equation}
Here $F_{\rm res}=F_{\rm swap}$ for the local swapped state and $F_{\rm res}=F_{\rm out}$ for the delivered pair \cite{Horodecki1999}. At fixed conditional phase and other noise parameters, the dependence on output-mode overlap is equivalent to an additional phase-flip channel:
\begin{equation}
 \mathcal T_M=\mathcal D_Z(M)\circ\mathcal T_1,\qquad
 \mathcal D_Z(M)(\sigma)=\frac{1+M}{2}\sigma+\frac{1-M}{2}Z\sigma Z.
\end{equation}
Thus $M=0.99$ adds a phase-flip probability of $0.5\%$. For $M=1$ at $T_*$, the local swapped state has $p_Z=0.0037$ and $p_X=p_Y=0.0160$.

For entanglement-based BB84, the $Z$ basis supplies the key and the $X$ basis estimates phase errors. The error rates and secret fraction are
\begin{equation}
 Q_Z=p_X+p_Y=\frac{1-c_z}{2},\qquad
 Q_X=p_Z+p_Y=\frac{1-c_\perp}{2},\qquad
 r=\big[1-h_2(Q_Z)-h_2(Q_X)\big]_+,
\end{equation}
where $h_2$ is the binary entropy and $[x]_+=\max(0,x)$ \cite{Shor2000}.

\begin{center}
\begin{tabular}{lrrrr}
\toprule
Resource delivered at $100$ km & $\overline f_{\rm tel}$ & $Q_Z$ (\%) & $Q_X$ (\%) & $K$ (bit/s)\\
\midrule
$M=1$, $\tau_p=200$ ps &0.9667&4.13&2.93&5.11\\
$M=1$, $\tau_p=250$ ps &0.9667&4.13&2.93&5.67\\
$M=0.99$, $\tau_p=250$ ps &0.9667&3.82&3.08&5.26\\
Conventional two-state beam splitter &0.9807&1.93&1.93&6.29\\
\bottomrule
\end{tabular}
\end{center}
These results use the fixed-$T_*$ operating points and memory cutoffs of the pair-rate comparison. Both output-mode coherence and storage time affect the errors. The beam-splitter reference gives the highest key rate for these parameters.

\section{Optimal interaction range}
Each incident photon has a Gaussian intensity envelope of rms width $\sigma$. The relative longitudinal coordinate has density
\begin{equation}
 p_0(r)=\frac{e^{-r^2/(4\sigma^2)}}{2\sqrt\pi\sigma},\qquad
 \langle r^2\rangle_0=2\sigma^2.
\end{equation}
We model the difference between the even- and odd-parity interaction energies by a Gaussian spatial profile:
\begin{equation}
 \delta(r)=V_e(r)-V_o(r)=
 \frac{g}{\sqrt{2\pi}a}e^{-r^2/(2a^2)},\qquad
 g=\int_{-\infty}^{\infty}\delta(r)\,dr.
 \label{eq:gaussiankernel}
\end{equation}
Here $a$ is the interaction range in the photons' relative coordinate, $\sigma$ is the incident packet width, and $g$ is the integrated coupling. The mean splitting at the input is
\begin{equation}
 \overline\Delta=\langle\delta\rangle_0
 =\frac{g}{\sqrt{2\pi}\sqrt{a^2+2\sigma^2}}.
 \label{eq:gaussianmean}
\end{equation}
At fixed $g$, increasing $a$ makes the conditional phase more uniform but reduces the mean precession rate.

In the dispersionless approximation, the parity-mode overlap is
\begin{equation}
 \Gamma(T)=\int_{-\infty}^{\infty}p_0(r)
 \exp[-i\delta(r)T/\hbar]\,dr.
 \label{eq:frozenoverlap}
\end{equation}
At fixed $a$ and $\sigma$, it depends on $gT$, so increasing $g$ shortens the time required to accumulate a given phase. Dispersion introduces a second timescale and breaks this rescaling. Related wavepacket effects in photon phase gates are discussed in Refs.~\cite{He2012,Brod2016}. The first two moments of the splitting give
\begin{equation}
 \frac{\langle\delta^2\rangle_0}{\overline\Delta^2}
 =\frac{a^2+2\sigma^2}{a\sqrt{a^2+4\sigma^2}},\qquad
 \epsilon^2\equiv
 \frac{\langle(\delta-\overline\Delta)^2\rangle_0}
 {\overline\Delta^2}
 =2\left(\frac{\sigma}{a}\right)^4
 +O[(\sigma/a)^6].
 \label{eq:gaussianvariance}
\end{equation}
At $T_q=\pi\hbar/(2\overline\Delta)$, the mean conditional phase is $\pi/2$. Expanding about this phase gives
\begin{equation}
 -\operatorname{Im}\Gamma(T_q)
 =1-\frac{\pi^2\epsilon^2}{8}+\cdots.
\end{equation}
With the Pauli corrections of Eq.~(\ref{eq:bellmap}) and equal spin-dephasing rates, the node fidelity is
\begin{equation}
 F_n(T)=\frac{1+\lambda^2-2\lambda\operatorname{Im}\Gamma(T)}4,
 \qquad \lambda=e^{-T/T_\phi}.
 \label{eq:gaussianfidelity}
\end{equation}
For $q=a/\sigma\gg1$ and $T_q/T_\phi\ll1$, its leading infidelity is
\begin{equation}
 1-F_n\approx\frac{T_q}{T_\phi}+\frac{\pi^2\sigma^4}{8a^4}
 \approx\zeta q+\frac{\pi^2}{8q^4},\qquad
 \zeta=\frac{\pi\hbar\sqrt{2\pi}\sigma}{2gT_\phi}.
 \label{eq:gaussianbalance}
\end{equation}
The two terms describe spin dephasing and the effect of phase variation across the wavepacket. Balancing them gives
\begin{equation}
 q_{\rm opt}\approx\left(\frac{\pi^2}{2\zeta}\right)^{1/5},
 \qquad
 (1-F_n)_{\min}\approx
 \frac54\left(\frac{\pi^2}{2}\right)^{1/5}\zeta^{4/5}.
 \label{eq:gaussianoptimum}
\end{equation}
The optimal range therefore grows as $g^{1/5}$ and the minimum node infidelity falls as $g^{-4/5}$ for fixed packet width and Markovian spin dephasing. Source and background errors give $F_{\rm swap}=1/4+A(F_n-1/4)$, which approaches $0.9893$ as $F_n\to1$ for the stated inputs. Photon loss enters the success probability and can shift the maximum distribution rate to a shorter interaction time.

\section{Two-photon propagation}
The two parity amplitudes obey
\begin{equation}
 i\hbar\partial_Tu_j(r,T)=
 [-\hbar\kappa\partial_r^2+V_j(r)]u_j(r,T),\qquad
 V_o=-0.1V_e,\quad V_e-V_o=\delta(r),
 \label{eq:gaussianpropagation}
\end{equation}
with initial amplitudes $u_j(r,0)=\sqrt{p_0(r)}$. The dispersion coefficient is $\kappa=0.8356\,\mu$m$^2$/ps; we use $\sigma=5\,\mu$m and $T_\phi=2$ ns. The overlap $\Gamma(T)=\int u_o^*(r,T)u_e(r,T)\,dr$ includes all arrival times. Spin dephasing enters through Eq.~(\ref{eq:gaussianfidelity}), while population decay reduces the pair survival probability.

Table~\ref{tab:gaussian} lists the maximum local fidelities from the range scan, with time refined near each peak. At these optical operating points, only the memory waiting-time cutoff is optimized using Eq.~(\ref{eq:renewal}). For $g=1000\,\ue\,\mu$m, the selected range is $a=17.55\,\mu$m and $\Gamma\approx0.055-0.984i$ at $T=48.1$ ps. The node fidelity is $F_n=0.9688$, and source and background errors give $F_{\rm swap}=0.959$. At $100$ km, the optimal cutoff is $D=114$ attempts and the mean delivered fidelity is $F_{\rm out}=0.95015$; the oldest accepted pair has fidelity $0.938$. For $\tau_p=250$ ps, the rate is $9.24$ pairs/s, exceeding the conventional beam-splitter rate by $6.6\%$ and the reference with partial acceptance of ambiguous outcomes by $1.9\%$. Reducing $\tau_p$ to $200$ ps gives $P_{\rm swap}=0.366$ and $R=8.39$ pairs/s. For $g=1400\,\ue\,\mu$m, $a=18.94\,\mu$m and $D=365$ meet the same fidelity requirement. Its rate at $200$ ps is $10.74$ pairs/s.
\begin{table}[ht]
\centering
\caption{Operating points for an interaction with a Gaussian spatial profile. The units of $g$, $\overline\Delta$, and $T$ are $\ue\,\mu$m, $\ue$, and ps. $F_{\rm swap}$ is the local swapping fidelity. Both $P_{\rm swap}$ and $R$ use $\tau_p=250$ ps. The rate $R$ is evaluated at $100$ km with the waiting-time cutoff optimized under the average-fidelity requirement $F_{\rm out}\geq0.9501$. A zero rate means that the fidelity requirement is not met.}
\label{tab:gaussian}
\begin{tabular}{rrrrrrrr}
\toprule
$g$ & $a/\sigma$ & $\overline\Delta$ & $T$ & $|\Gamma|$ &
$F_{\rm swap}$ & $P_{\rm swap}$ & $R$ (s$^{-1}$)\\
\midrule
354.63 &2.760&9.12&109.7&0.957&0.908&0.246&0\\
700 &3.235&15.82&64.0&0.980&0.947&0.355&0\\
1000 &3.510&21.08&48.1&0.986&0.959&0.403&9.24\\
1400 &3.787&27.63&36.8&0.990&0.966&0.441&11.56\\
\bottomrule
\end{tabular}
\end{table}

For the dispersionless calculation, we set $r=2\sigma x$. A logarithmic scan over $g=300$--$10^5\,\ue\,\mu$m, with range optimization at $T_q$, gives an error exponent of $-0.799$ from the last ten points. Dispersive propagation at the same ranges and times for $g=3000,10^4,3\times10^4,10^5\,\ue\,\mu$m gives $-0.800$, consistent with the $g^{-4/5}$ asymptote. Propagation calculation uses a split-step Fourier method on a $512\,\mu$m domain with 4096 points and a $0.1$-ps time step. We sample $0\leq T\leq200$ ps every $0.5$ ps and refine selected peaks on a $0.05$-ps grid using a $1024\,\mu$m domain and a $0.05$-ps evolution step.

\section{Waveguide geometry and polarization control}
The proposed device is a planar GaAs-based semiconductor waveguide containing quantum wells, with two polarization components of the lower-polariton branch. Input and output dielectric gratings define a single-pass propagation length $\ell$. The input gratings resonantly populate two distinguishable modes $c,t$ with overlapping transverse profiles. Mode-selective output coupling preserves their labels for independent polarization readout.

For narrow-band packets with matched group velocity,
\begin{equation}
 T=\frac{\ell}{v_{\rm gr}},\qquad
 v_{\rm gr}=\frac{\partial\omega_{\rm LP}}{\partial k}.
 \label{eq:gratingtime}
\end{equation}
The injection frequency and wave vector select the operating point on the dispersion. At $v_{\rm gr}=1.67\,\mu$m/ps, the uniform-interaction time $T_*=51.7$ ps corresponds to $\ell=86.3\,\mu$m. The spatially dependent operating point $T=48.1$ ps corresponds to $\ell=80.3\,\mu$m. Maintaining packet overlap requires $|v_c-v_t|T\ll\sigma$ and an injection delay much shorter than $\sigma/v_{\rm gr}=3.0$ ps for $\sigma=5\,\mu$m. The operating points in the two guided modes must be chosen to match their group velocities. Polarized propagation has been demonstrated in Ref.~\cite{Walker2015}. The proposed gate requires coherent transmission of both polarization components throughout the guide. For the spin Hamiltonian used here, a residual linear-polarization splitting $\delta_{\rm lin}$ must obey $|\delta_{\rm lin}|T/\hbar\ll1$. Polarization-dependent transmission and scattering out of the selected modes must also be small. The coupling parameter $\eta_w$ includes injection and extraction for each photon. The propagation length $\ell$ controls the interaction duration, while $a$ describes the interaction range in the relative coordinate of the two photons.

\section{Polariton parameters and the short-range limit}
To compare with a short-range interaction, we use
\begin{equation}
 i\hbar\partial_Tu_j=
 [-\hbar\kappa\partial_r^2+g_jf_a(r)]u_j,\qquad
 f_a(r)=\frac{e^{-r^2/(2a^2)}}{\sqrt{2\pi}a},
 \quad r=v(\tau_c-\tau_t),\quad T=z/v.
\end{equation}
Here $\tau_c,\tau_t$ are the photon arrival times at a common propagation plane, $v$ is their matched group velocity, and the amplitudes are normalized to photon flux. Both modes have the same second-order dispersion. The range $a$ sets the interaction's dependence on relative arrival time; the propagation length is $vT$. For $\sigma=5\,\mu$m, $\kappa=0.8356\,\mu$m$^2$/ps, $a=0.2\,\mu$m, and $g_o=-0.1g_e$, an input mean splitting of $20\,\ue$ requires $g_e-g_o=354.63\,\ue\,\mu$m. With the same source and spin noise, propagation over $0\leq T\leq200$ ps gives a maximum swapping fidelity of $0.756$ when all arrival times are accepted. Halving $a$ at fixed integrated coupling gives $0.757$. The first row of Table~\ref{tab:gaussian} keeps this coupling and increases the range to $13.8\,\mu$m. The mean splitting falls to $9.12\,\ue$, while the more uniform phase raises the maximum fidelity to $0.908$. Maintaining a mean splitting of $20\,\ue$ at a larger range requires a larger $g$, as given by Eq.~(\ref{eq:gaussianmean}). Lifetimes of $180\pm10$ ps have been measured in planar cavities \cite{Steger2015}, while single-polariton propagation \cite{SuarezForero2020} and guided polariton transport \cite{Rozas2024} have also been demonstrated. These demonstrations provide an experimental basis for exploring conditional Larmor precession and entanglement swapping in GaAs-based microcavity waveguide.

\bibliographystyle{apsrev4-2}
\bibliography{references}